%Paper: 9203066
%From: VAFA@huhepl.harvard.edu
%Date: Tue, 24 Mar 1992 17:16:33 -0500 (EST)

\input harvmac
\Title{\vbox{\baselineskip12pt\hbox{HUTP-92/A013}\hbox{SISSA 44/92/EP}}}
{\vbox{\centerline{Massive Orbifolds }}}
\vglue 1cm
\centerline{Sergio Cecotti}
\centerline{International School for Advanced Studies, SISSA-ISAS}
\centerline{Trieste and I.N.F.N., sez. di Trieste, Trieste, Italy}
\vglue 1cm
\centerline{{Cumrun Vafa}}
\centerline{Lyman Laboratory of Physics}
\centerline{Harvard University}
\centerline{Cambridge, Ma 02138, USA}
\vglue 2cm

We study some aspects of 2d supersymmetric sigma models on orbifolds.
It turns out that independently of whether the 2d QFT is conformal
the operator products of twist operators are non-singular, suggesting
that massive (non-conformal) orbifolds also `resolve singularities' just as in
the conformal case.  Moreover we recover the OPE of twist operators for
conformal theories by considering the
UV limit of the massive orbifold correlation functions.   Alternatively,
we can use the OPE of twist fields at the conformal point to derive
conditions for the existence of non-singular solutions to special
non-linear differential equations (such as Painleve III).
%\draft
\Date{3/92}

Orbifolds \ref\orb{L.Dixon, J. Harvey, C. Vafa and
E. Witten, Nucl. Phys. B274 (1986) 285.}\ are among the simplest class of
solutions to string theory.
The simplest type of orbifolds arise by considering
propagation of strings on quotients of tori by isometries.  In such
cases one generically ends up with a quotient which is singular due
to the existence of fixed points of the group action.  However
it turns out that despite the singularity of the geometry, the 2d CFT
corresponding to it is non-singular; this simply means that the correlation
functions of these theories, which can be computed in closed form
\ref\hv{S.Hamidi and C. Vafa, Nucl. Phys. B279 (1987) 465.}\ref\dix{L. Dixon,
D. Friedan, E. Martinec and S. Shenker, Nucl. Phys. B282
(1987) 13.}\ref\ber{M. Bershadsky and A. Radul, Int. J. Mod. Phys. A2 (1987)
165.}\ref\Kn{V.G.Knizhnik, Comm. Math. Phys.112 (1987) 567.}\ are finite.  The
main correlation functions to check are those of twist fields which create
strings which are stuck at the vicinity of the fixed points.
Let us consider strings propagating on a Kahler manifold $M$ of dimension
$d$.  Let $G$
be a discrete group of isometries of $M$.  Then $M/G$ is the orbifold
we wish to study.  If $M$ is Calabi-Yau and $G$ leaves the holomorphic
$d$-form invariant, then $M/G$ is itself a (possibly singular) Calabi-Yau
manifold.  However if $M$ is not a Calabi-Yau manifold then the resulting
QFT is not conformal and will have massive excitations\foot{In order to have an
asymptotically free theory we assume that the first chern class of $M$
evaluated on any two cycle is positive.}.  We call the
resulting orbifolds {\it massive orbifolds}.  The case of interest for us is
the case of superstrings which corresponds
to supersymmetric sigma models on $M/G$.

Many of the properties of orbifolds discussed in the context of
conformal theories \orb\ continue to hold in the massive case as well.
In particular the Hilbert space of strings on $M/G$ differs from that of $M$ by
introducing one twisted sector $H_g$ for each element $g\in G$ which correspond
to strings which are periodic up to $g$ action, and then projecting to the
$G-$invariant subsector.  In order to study potential singularities
of orbifold correlation functions we need to concentrate on fields
which are stuck at the fixed points of the $G$-action.
These
states will correspond, for each $H_g$ (in the
Ramond sector), to the lightest states in
that sector.  Luckily, these lightest states in the context
of superstrings will have topological significance.  In particular
the ground states in $H_g$ are in one to one
correspondence with the cohomology elements of $M_g$ where $M_g\subset M$
denotes the submanifold of $M$ left invariant by $g$.  This follows from
the topological nature of ground states of sigma models \ref\wit{
E. Witten, Nucl. Phys. B202 (1982) 253.}\ applied
to the $g-$th twisted sector \orb .  They are also in one to one correspondence
with the chiral fields in the NS sector of $H_g$ by spectral flow.
These chiral twist fields are what we will mainly study.
It is well known that the chiral fields
form a ring called the chiral ring which has topological properties.
This ring has been studied in various works \ref\chir{W. Lerche,
C. Vafa and N. Warner, Nucl. Phys. B324 (1989) 427.}\ref\mar{E. Martinec, {\it
Criticality, Catastrophe and Compactifications},
V.G. Knizhnik memorial volume, 1989.
}\ref\cecgi{S. Cecotti, L. Girardello and A. Pasquinucci, Nucl. Phys.
B328 (1989) 701; Int. J. Mod. Phys. A6 (1991) 2427.}.
  To find the chiral fields of the full
theory we just need to take a direct sum of chiral fields from each
twisted sector and project the total Hilbert space to the $G-$invariant
subspace.

The chiral fields in the untwisted sector $H_1$ form
a ring which is a quantum deformation (due to instantons) of the cohomology
ring of $M$ \ref\witt{E. Witten, Comm. Math. Phys. 118
(1988) 411; Nucl. Phys. B340 (1990) 281.}.  As we will see below a simple
modification of this approach allows us to compute the chiral ring
for the $M/G$ theory.  So it is natural to expect that this ring
should be a quantum deformation of the cohomology ring of $M/G$.  Such an
expectation is indeed justifiable in the context of Calabi-Yau manifold where
one can
resolve the singularities
and obtain a non-singular Calabi-Yau manifold.  Since for non-singular
Calabi-Yau manifolds this is indeed true, and since orbifolds are physically
non-singular limits of them, the same is true for the chiral ring
for the orbifolds.   In particular studying the structure of $M_g$
allows one to compute the Euler character of $M/G$ rather easily \orb \
which agrees with the corresponding computation for the resolved
Calabi-Yau manifold \ref\roan{S.S. Roan, Int. J. of Math., v.1 (1990)
211.}.  This correspondence between the chiral
ring of $M/G$ theory and the (quantum) cohomology
ring of the resolved manifold $M/G$ (even ignoring the ambiguities
in the resolution) turns out not to continue beyond
the Calabi-Yau case.  Even the dimensions of the two rings in general will
be different.  In a sense in the non-Calabi-Yau case
we get elements in the chiral ring
of the orbifold which morally speaking are the analogs of `harmonic
forms with fractional degree', and so have no simple geometrical
analog on the manifold\foot{They are cohomology elements on the {\it
loop space} of the orbifold.}.

Instead of being general it is more instructive to consider a concrete
example:  Take $S^2$ with a round metric and  let $G$ be the $Z_n$ group
generated by $g$ corresponding to rotation of $S^2$ about an axis by $2\pi/n$.
Let us count the ground states in each sector.  In the untwisted sector
we have the ordinary cohomology elements of $S^2$ which we represent
by $1,x$ where $x$ is the two form.  In the sector $H_{g^l}$
for $1\leq l \leq n-1$ we get two fixed points corresponding to
the `north' (N) and `south' (S) poles of $S^2$, and the chiral fields
corresponding to the north pole in the $l$-th sector we denote by $N_l$ and
to the south pole in the $(n-l)$-th sector\foot{Note that the reason for change
of convention
for the north versus south pole is that holomorphicity
of strings on the north pole and on the south pole correspond
to twisted states with opposite twists.}
we denote by $S_l$.  All of these elements are $G$
invariant and so they survive the G-projection.  Now it is clear
that these elements are not related to the cohomology elements of the
resolution of $S^2/G $ which is itself an $S^2$.  Nevertheless
it is not too difficult to compute the ring they generate.  The
way to do this is to use the topological formulation of sigma
models \witt\ and generalize it to the orbifold case.
As reviewed at length in \ref\cv{
S. Cecotti and C. Vafa,  Nucl. Phys. B367 (1991) 359.}\
the topological correlation functions encode the full
information about the chiral ring of the non-topological
theory, and in particular the ring can
be read off from the two and three point functions of the topological theory.
 For the ordinary sigma models we represent cohomology elements by cycles of
dual degree.
The topological two point function of these elements is just
the intersection number of the corresponding cycles and the three
point functions of the chiral fields, which using the topological metric (two
 point function) defines
the ring, is given by \witt\
\eqn\ring{C_{ijk}=<\phi_i (z_1) \phi_j(z_2) \phi_k(z_3)>=\sum_{inst.}exp(-S)}
where sum is over holomorphic instantons which map $CP^1$ holomorphically
to $M$ subject to sending $z_1$ to the $C_i$ cycle, $z_2$ to the $C_j$
cycle and $z_3$ to the $C_k$ cycle. The instantons contribute
if there is a discrete number of them\foot{More precisely, if the
virtual dimension of the relevant moduli space vanishes. In the general
case the instanton number should be
 replaced by the Euler character of a certain
bundle \ref\general{E. Witten, {\it Mirror manifolds and topological field
theory}, Princeton preprint IASSNS-HEP-91/83 (1991).}.} (if there is a family
of them it is defined to be zero). Here $S$ denotes the instanton action.
Applying these to the sphere we have the two and three point functions
$$<1 x>=1$$
$$<x x x >=\beta$$
where $\beta$ denotes the exponential of degree one instanton action (which in
this
case is (up to a phase) just the exponential of the minus the area of the two
sphere).
The last identity follows from the above rules and noting that $x$ is
represented as a 0-cycle, i.e., any point on $S^2$.
{}From these one reads the ring relation
$$x^2=\beta$$
Now we come to the orbifold case.  The first thing to note is that
we have a selection rule for interactions of strings on orbifolds:
 A correlation involving strings in $(g_1,...g_k)$
sectors is non-vanishing only if $\prod g_i =1$.
{}From the above selection rule we see that the two point function does not
vanish only between sectors which are inverse of one another.  Consider the
Hilbert spaces $H_g$ and $H_{g^{-1}}$.  Then we represent the
chiral states in these sectors by cycles of $M_g$.  Note that
$M_g=M_{g^{-1}}$.  Then the two point functions of these
chiral fields is simply the intersection of corresponding cycles
of $M_g$.  In the case of $S^2/Z_n$ we have
$$<N_l N_{n-l}>=<S_l S_{n-l}>=1$$
and the rest of the two point functions (of twisted chiral fields) vanish.
All we need to do to complete the picture is to generalize \ring\
for the orbifold case.  The generalization follows form the rules
of computation of correlation functions for orbifolds and leads to the
following
 answer \ref\Zas{In the topological setup this is currently being investigated
by
 E. Zaslow.}:
Consider a Riemann surface $\Sigma$ such that $\Sigma /G$ is a two sphere
with three branches points at $z_1,z_2$ and $z_3$.  Moreover, as we go around
these points on the two sphere we get monodromies of the fiber given
by monodromy elements $g_1,g_2,g_3$.  Now fix three cycles in $M_{g_1}$,
$M_{g_2}$ and $M_{g_3}$ corresponding to elements of chiral ring
whose correlation functions we wish to compute.  Then we just use
the same formula as \ring\ except that the space of instantons
is  holomorphic maps from $\Sigma$ to $M$, such that it is `G-equivariant'
which means that the points on $\Sigma$ which are $g$-transform of
one another, get mapped to points of $M$ which are $g$-transform of
each other and such that the points corresponding to the pre-images
of $z_1,z_2,z_3$ in $\Sigma$ get mapped to the corresponding cycles in
$M_{g_1},M_{g_2},M_{g_3}$ respectively.
Also the instanton action is the one corresponding to the map from
the sphere, which is the same as the action for the map from $\Sigma$
to $M$ divided by the order of the group $S=S_{\Sigma }/|G|$.

Going back to our concrete example of $S^2/Z_n$ we compute
the non-trivial three point functions and find
\eqn\simp{<N_{l_1}N_{l_2}N_{n-l_1-l_2}>=<S_{l_1}S_{l_2}S_{n-l_1-l_2}>=1}
\eqn\ssim{<N_l S_l\ x>=\beta ^{l/n}}
(with $l_1+l_2<n$).
The equation \simp\ and the two point functions imply that
$$N_{l_1}N_{l_2}=N_{l_1+l_2}=N^{l_1+l_2} \qquad and \quad (N\rightarrow S)$$
\eqn\rsimr{N^n=S^n=x}
where $N$ and $S$ denote the corresponding chiral fields in the
singly twisted sectors. These ring relations are more or less
obvious when one pictures an $l$-th twisted string state as a string
wrapped on a small arc about a pole which spans $2\pi l/n$ in angle.
 The equation \ssim\ might look a little less obvious, but
that can also be easily understood when we realize that the field
$x$ is represented by a point on the sphere
and so \ssim\ corresponds to a holomorphic map which maps
the worldsheet sphere to the fraction $l/n$ of the target sphere, mapping
the north pole (where we have inserted $N_l$)
to north pole, south pole (where we have inserted $S_l$) to south pole, and the
third point (where we have inserted $x$) to a point on the sphere corresponding
to $x$.  Clearly
the action for this is $l/n$ times the action of an instanton of
the original theory, which thus explains the factor $\beta^{l/n}$
in \ssim .  From this equation we can read the ring relations
\eqn\rrel{N^l x=\beta^{l/n} S^{n-l} \qquad and \quad
(N\leftrightarrow S)}
\eqn\mrel{N^l S^l=\beta ^{l/n}}
Note that these relations are compatible with the original ring
relation $x^2= \beta$, as would follow by multiplying \rrel\
by $x$. In particular $x$ forms a two dimensional representation
when acting on the vacuum states corresponding to ($|N^l \rangle$,
$|S^{n-l} \rangle $).

Note that given the fact that $x$ is a $(1,1)$ form, equation \rsimr\
shows that $N,S$ are $(1/n,1/n)$ forms.  This statement can be made precise
in the UV limit (i.e., when instantons are suppressed
and chiral fermion number is conserved) by showing that $S,N$
have left-right chiral fermion numbers given by $1/n$.  This shows that we
cannot hope to have a simple geometrical meaning attached to these fields
in general as mentioned earlier.

 To see whether or not the twist fields correspond to singular
fields, given the fact that they have a very simple  and non-singular ring
relation \rsimr\ all we have to check is whether their
normalizations are singular.  For that purpose it suffices to consider
the Ramond ground states they create when acting on vacuum.  So we need to
compute
$$ e^{q_l}=\langle \bar N ^l|N^l \rangle = \langle \bar S^l |S^l
\rangle$$
$$e^{p_l}=\langle \bar N^l |S^{n-l}\rangle =\langle \bar S^l|N^{n-l}\rangle $$
where the second equality in the above equations simply follow from
the symmetry in exchanging the north and south poles,
and the other norms between twist fields which we have
not written down are identically zero due to the $Z_n$ symmetry
of orbifolds mentioned before.  These norms are in general function
of the area $A$ of the sphere which here is encoded in $\beta$
by $A= -\rm log |\beta |$ .  So
we wish to find the $\beta $ dependence of the above metrics, and see
whether or not they are singular for any value of $\beta$.
The machinery for doing this has been discussed in \cv .
In particular, the case corresponding to the un-orbifoldized $S^2$,
leading to the computation of $\langle \bar x | x\rangle$ and $\langle
\bar 1| 1\rangle $
has been discussed in \ref\cvb{S. Cecotti and C. Vafa, Phys. Rev. Lett. 68
(1992) 903.}.
 We will now summarize the results of \cv\ which are relevant
for us here.  The main conclusion is that if we consider
perturbing an $N=2$ theory by  perturbing the superpotential
$$ \Delta L =\int \Delta  W( \phi , t) d^2zd^2\theta + c.c.$$
where $t$ denotes coupling constants,
and if we denote the action of the chiral field $dW$
(the variation of $W$ with respect to $t$) on the ground
states by the matrix $C$, and if we denote the metric of the
ground states by the matrix $g$, then (in a holomorphic basis) we have
\eqn\main{\bar \partial (g \partial g^{-1})=[C,gC^{\dagger }g^{-1}]}
where partials refer to derivatives with respect to $t$ and $\bar t$.
In addition if the two point function of the chiral fields is denoted
by $\eta$, then $CPT$ invariance of the theory implies
\eqn\real{(g^{-1}\eta)(g^{-1} \eta)^*=1}
Now we apply these to our case where $W=-{\rm log }\beta \ x$
and $t=\beta$, concentrating on the metric for the twisted ground
state.  Using the form of $\eta$ discussed above, and applying
\real\ we find that $e^{p_l}=0$ and so we have only $e^{q_l}$ to
compute.  In this case $g$ is a diagonal matrix in the basis given
by powers of $N$ and $S$.  The matrix $C$ can be represented
using the ring relations as
$$C={-1\over \beta}\left(\matrix {0&\beta^{l\over n}\cr \beta^{l-n\over
n}&0\cr}
 \right)$$
acting on the vacuum subspace spanned by $(|N^l \rangle ,|S^{n-l} \rangle )$.
Using this, and defining
\eqn\chv{{\tilde q_l}=2q_l-{2l-n\over n}{\rm log} |\beta | \qquad
z=2\beta^{1/2}
}
we find that \main\ reduces to the sinh-Gordon equation
\eqn\sinc{\partial \bar \partial {\tilde q_l}=4\ {\rm sinh}\ \tilde q_l}
where derivatives are in terms of $z,\bar z$.
One can also show that by fermion number conservation $\tilde q_l$
is only a function of $|z|$, which means we are looking for rotationally
symmetric solutions for sinh-Gordon equation.  In this case, this
equation reduces to a special case of Painleve III equation \ref\pain{
B.M. McCoy, C.A. Tracy and T.T. Wu, J. Math. Phys. 18 (1977) 1058; and
E. Barouch, Phys. Rev. B13 (1976) 316\semi
A.R.Its and V.Yu.Novokshenov, {\it The isomonodromic Deformation
Method in the Theory of Painleve Equations}, Lecture Notes in Mathematics
1191, Springer-Verlag, Berlin 1986.}.
We are looking for solutions of this equation which give finite
values for $q_l$.  This in particular means, using \chv\ that as $|\beta|
\rightarrow 0$, we must have
$${\tilde q_l}\rightarrow -{2(2l-n)\over n}{\rm log} |z|/2 + const. \qquad
z \rightarrow 0$$
It turns out that there are regular solutions to Painleve equation for all
$z$, with this boundary conditions, if and only if the small $|z|$
behavior is given by \pain\
$$\tilde q_l \simeq
r {\rm log}|z|+s ; \qquad {\rm with}\qquad e^{s/2}={1\over 2^r}
{\Gamma ({1\over 2}-{r\over 4})\over \Gamma ({1\over 2}+{r\over 4})}$$
This is consistent with our choosing $r=-2(2l-n)/n$.
So we have learnt that there exists a regular solugion for $q_l$ for all
$\beta$;
this includes the region  $|\beta |>1$ (which corresponds to super-strong
couplings--the sphere with {\it negative} area).  Moreover we deduce
that in the limit that the area of the sphere goes to infinity
we have
\eqn\qcon{e^{q_l(0)}=2^{2l-n\over n}{\Gamma ({l\over n})\over \Gamma (1-{l\over
n})}}
This can be compared with the OPE of the conformal field theory,
because as the area of the sphere goes to infinity we have effectively
a conformal theory on the plane modded out by $Z_n$ (the south pole
can be taken to be at infinity).  To compare the OPE of the twist
fields with the above result we will have to normalize our fields
properly.  This is easily done as follows:  Let ${\tilde N_{l}}$ denote
the twist field of the $l-$th sector
creating a normalized ground state and ${\tilde x}$ denote
the normalized field corresponding to the Kahler form.  Then we must have
in the UV limit
$${\tilde N_l}=\left[ {\langle \bar 1| 1\rangle \over \langle
\bar N^l |N^l \rangle }\right]^{1/2} N^l=(2A)^{1/2}2^{-{l\over n}+{1\over 2}}
\left[ {\Gamma ({n-l\over n})\over \Gamma ({l\over n})}\right]^{1/2}N^l$$
$${\tilde x}=\left[{\langle \bar 1| 1\rangle \over \langle \bar x |x\rangle }
\right]^{1/2}x =(2 A)\ x$$
where we have used that in this limit $\langle \bar 1|1\rangle =\langle
\bar x |x \rangle^{-1}=(2A)$.
{}From these and the ring relation $N^n=x$ we can compute the normalized OPE's:
$${\tilde N}_{l_1}{\tilde N}_{l_2}{\tilde N}_{l_3}=(4 A)^{1/2}
\left[ \prod_i {\Gamma ({n-l_i\over n})\over \Gamma ({l_i\over n})}
\right]^{1/2}{\tilde x}$$
with $\sum l_i =n$, and similarly with $l_i \rightarrow n-l_i$ if $\sum
l_i=2n$.
  These results are in perfect agreement with the computations
at the conformal point  \dix \ber \Kn\ using very
different methods from the above (compare in particular with eq. (7.41)
in \Kn )\foot{
Note that even though we have phrased our computations in the supersymmetric
set up, our result for the OPE of twist fields is correct also for the
purely bosonic one, because the twist fields corresponding to the
fermionic degrees of freedom have OPE equal to one, as follows from
bosonizing them and representing them as exponential of bosonic fields.
Thus the fermionic degrees of freedom on the left cancel $\tilde x$
on the right, which is purely fermionic, in the OPE written above.}. It is very
amusing that consistency
of the {\it massive orbifold} correlations imply the correct OPE
for the conformal theory computation.  This result is similar to what
was found in many examples considered in \cv .
 The above agreement with CFT computation is in our opinion a strong
check for the self-consistency of massive orbifold QFT.
We can also reverse the logic:  Assuming that the massive orbifold
theory is consistent, we can view the computation of the twist
correlations in the conformal
theory as solving for the good boundary
conditions which lead to non-singular solutions for certain highly non-trivial
non-linear differential equations such as Painleve III.  In fact
(combined with the result of our previous paper \cvb ), in this way we have
recovered {\it all} the boundary conditions which give regular
solutions to Painleve III by considering supersymmetric sigma models on
$CP^1$ and its orbifolds.  In a similar way by considering various orbifolds
of $CP^n$, we can find radial solutions for affine Toda $A_n$.
It would be interesting to see if this leads to all radial solutions of
affine toda with all consistent boundary conditions.
It would also be interesting to study the OPE of twist correlators for
non-abelian orbifolds.  This is a subject which has not been extensively
studied even in the conformal case (see however \ref\gat{B. Gato,
Nucl. Phys. B334 (1990) 414.})
and it would be interesting to see if the above approach
allows one to compute their OPE as a regularity condition
for the existence of solution to some differential equations.
The self-consistency that we have found
for the QFT of massive orbifold sigma models is in harmony with
the result of \ref\gife{P.Fendley and
P.Ginsparg, Nucl. Phys. B324 (1989) 549.}\ where it was found
that orbifolds of non-critical statistical mechanical models are
also well behaved.

What physical lessons are we learning from this exercise?  For one thing
we learn that the singularities of geometry, which would also be singularities
of field theories (maps of world-line to the manifolds) disappear when
we consider `string theories' (i.e., maps of the world-sheet to the manifold).
The physical reason for this is the fact that strings are extended objects
and are less sensitive to singularities that would prove fatal
for point particle theories.  We have now seen that it is {\it not} because
of conformal invariance that these singularities disappear.  It
holds true even in the non-conformal case that we have studied in this
paper.

A second lesson we should learn is that to study the singularities
of $N=2$ supersymmetric quantum field theories, it seems to be sufficient
to concentrate on the chiral ring of the theory.  This is so in our example
because the twist fields, which were the potential source of singularities
are part of the chiral ring and can be studied using powerful
pseudo-topological
 techniques.  Thus the presence or absence of singularities
in the full quantum field theory can be deduced from the presence or absence
of singularities in the chiral ring itself.  This is not to say that
singularities never happen for $N=2$ supersymmetric quantum field theories.
For example if we consider the Landau-Ginzburg theory given by the
superpotential
$$W=(x^3+y^3+z^3-a \ xyz)$$
at $a=3$ it is singular (the dimension of the ring is infinite at this point).
This singularity in the ring is reflected in geometry of this theory as
(after a $Z_3$ twist) it corresponds to strings propagating on a
torus given by setting $W=0$ in $CP^2$ and at $a=3$ it corresponds
to a torus with a degenerate complex structure which is a singular
conformal theory (in the sense of having singular OPE's).
So the moral of the story is that the chiral ring encodes complete
information of whether the theory is singular.  Similar ideas
about the topological nature of singularities has previously been expressed in
the context of singularities of the two dimensional
black hole by Witten \ref\wittsn{E. Witten, Phys. Rev. D44 (1991) 314.}.
Moreover, in the supersymmetric version of the black hole Eguchi has found
a remarkable correspondence between the elements of the chiral ring and the
singular points of space-time geometry \ref\Egu{T. Eguchi, {\it
Topological field theories and the space-time singularity}, Univ. of Chicago,
preprint 1991.}\ just as in the orbifold case.
  So to find out whether
black hole singularity is resolved by the world sheet theory, we
should simply study the properties of this ring, similar to what
we have done here for $CP^1$ orbifolds,
and see whether the chiral ring elements have a non-singular normalized
OPE.

 The research of C.V. was supported in part by Packard Foundation
and NSF grants PHY-89-57162 and PHY-87-14654.

\listrefs

 \end